\documentclass[aps,preprint]{revtex4}%
\usepackage{amsfonts}
\usepackage{amsmath}
\usepackage{amssymb}
\usepackage{graphicx}%
\providecommand{\U}[1]{\protect\rule{.1in}{.1in}}

\begin{document}
\title[ ]{Conflict Between Classical Mechanics and Electromagnetism: The Harmonic
Oscillator in Equilibrium with a Bath}
\author{Timothy H. Boyer}
\affiliation{Department of Physics, City College of the City University of New York, New
York, New York 10031}

\begin{abstract}
It is pointed out that an electric charge oscillating in a one-dimensional
purely-harmonic potential is in detailed balance at its harmonics with a
radiation bath whose energy $U_{rad}$\ per normal mode is linear in frequency
$\omega$, $U_{rad}=const\times\omega,$ and hence is Lorentz invariant, as
seems appropriate for relativistic electromagnetism. \ The oscillating charge
is \textit{not} in equilibrium with the Rayleigh-Jeans spectrum which arises
from energy-sharing equipartition ideas which are valid only in
nonrelativistic mechanics. \ Here we explore the contrasting behavior of
harmonic oscillators connected to baths in classical mechanics and
electromagnetism. It is emphasized that modern physics text are in error in
suggesting that the Rayleigh-Jeans spectrum corresponds\ to the equilibrium
spectrum of random classical radiation, and in ignoring Lorentz-invariant
classical zero-point radiation which is indeed a classical equilibrium
spectrum. \ 

\end{abstract}
\maketitle

\section{Introduction}

\subsection{Harmonic Oscillators as Illustrative of the
Mechanics-Electomagnetism Conflict}

Ask a physicist to name the spectrum of random classical radiation with which
an oscillating classical charge in a one-dimensional purely-harmonic potential
will come to equilibrium at its harmonics, and, influenced by current modern
physics textbooks, he will invariably give the wrong answer. \ He will suggest
\textquotedblleft the Rayleigh-Jeans spectrum.\textquotedblright\ \ But this
suggestion is untrue. \ The correct answer is \textquotedblleft a
Lorentz-invariant spectrum.\textquotedblright\ The Rayleigh-Jeans spectrum is
associated with the energy-sharing equipartition ideas which are valid only in
nonrelativistic classical mechanics. \ On the other hand, electromagnetism is
a relativistic theory, and a classical charge oscillating in one dimension is
in equilibrium at its harmonics with a relativistically-invariant spectrum of
random radiation. \ In this article, we explore the conflict between Newtonian
mechanics and electromagnetism in connection with harmonic oscillators and
equilibrium baths.

The mismatch between Newtonian mechanics and classical electromagnetism goes
unappreciated by many physics students and by their instructors. \ The
mismatch is fundamental since Newtonian mechanics satisfies Galilean
invariance while electrodynamics is relativistically invariant. \ Yet the
mismatch is usually ignored. \ The mismatch is usually ignored in our classes
because relativity is treated as a specialty subject covered at the beginning
of a class on modern physics; only point collisions between particles are
considered for relativistic interactions, and relativity is regarded as needed
only in nuclear and elementary-particle physics. \ It is usually assumed that
relativity becomes significant only for situations out of our ordinary
experience when particles are traveling at velocities close to the speed of
light. \ Here we show that relativity is central to striking contrasts in
harmonic oscillator systems involving low particle velocities in conjunction
with ambient baths. \ The simple analysis exposes \ the profound conflict
between nonrelativistic mechanical systems and relativistic electromagnetic
systems. \ 

The essential contrast is as follows. \ A frictionless harmonic oscillator
system in Newtonian mechanics can oscillate forever with a constant amplitude
and energy. \ There is no need for any bath for the oscillator. \ However, if
the mechanical oscillator is connected by point collisions to a bath of
particles, the total system energy is shared between the oscillator and the
particle bath so as to give energy equipartition. \ In contradiction with this
mechanical situation, an electromagnetic oscillator involving the oscillation
of a charged particle cannot exist on its own; the charged particle must be
coupled to radiation. \ Within classical electrodynamics, there is no such
thing as an oscillating charge which exists without radiation. \ It turns out
that the spectrum of radiation which must provide the equilibrium bath for an
oscillating charge in a one-dimensional purely-harmonic potential is uniquely
defined by electromagnetic theory, and turns out to correspond to a
Lorentz-invariant spectrum of classical electromagnetic radiation. \ 

This last statement will come as a surprise to the many physicists who are
aware of the historical situation involving Planck's linear oscillator taken
in dipole approximation which will match the energy of the radiation bath at
the oscillator frequency, but will not determine the radiation spectrum.
\ Such an oscillator appears in many modern physics texts in connection with
the blackbody radiation spectrum.\cite{mod} \ However, any electrodynamic
oscillator must involve an oscillating charge with a finite non-zero amplitude
of motion in the harmonic potential. \ The moment the amplitude of oscillation
is non-zero, the charge radiates not only at the fundamental frequency (as
given by the dipole approximation), but also at all the harmonics. \ The
radiation at the harmonics determines the spectrum of the ambient radiation
whose absorption is needed to balance the emitted radiation. \ Thus by
considering the harmonics, one arrives at a unique spectrum for the
equilibrium bath for the oscillating charge. \ A recent relativistic
calculation shows\cite{Equ} that classical electromagnetic theory for the
one-dimensional electromagnetic harmonic oscillator uniquely determines the
radiation spectrum with which the oscillator is in equilibrium, and the
spectrum is that of Lorentz-invariance where the energy of a radiation normal
mode is proportional to the frequency of the mode, $U_{rad}=const\times\omega
$. \ The well-known radiation spectrum $U_{rad}(\omega)=(1/2)\hbar\omega$ of
classical electromagnetic zero-point radiation fits this Lorentz-invariance
requirement. \ 

\subsection{Outline of the Article}

The outline of the presentation is as follows. \ First we describe the assumed
mechanical and electromagnetic oscillators and their baths. \ Then we consider
the equilibrium situations for the oscillators. \ For the electromagnetic
oscillator, we first describe the dipole harmonic approximation which is of
historical importance in suggesting the Rayleigh-Jeans spectrum, and also the
nonlinear oscillators taken in the dipole approximation which seem to confirm
the Rayleigh-Jeans conclusion. \ Then we sketch the extension of the
oscillator analysis to quadrupole radiation at double the frequency of the
oscillator and note that equilibrium requires a linear (and hence
relativistic) radiation spectrum for equilibrium. \ Next we turn to adiabatic
changes of the oscillator's fundamental frequency, and see how these changes
fit with thermodynamic ideas of the equilibrium temperature. \ The final
analysis notes the limitations in the use of a one-dimensional purely-harmonic
oscillator. \ Finally, we give some concluding remarks. \ 

\section{The Oscillators}

\subsection{Mechanical Oscillator}

Here we discuss the conflict between Newtonian mechanics and electromagnetism
in connection with equilibrium for small harmonic oscillators in a bath. \ In
the case of classical mechanics, the oscillator consists of a particle
attached to a spring along the $x$-axis in the absence of any friction. \ The
particle of mass $M$ is attached to a spring of constant $\kappa$ so that
Newton's second law gives the equation of motion due to the linear restoring
force $\mathbf{F}=-\widehat{x}\kappa x$ as%
\begin{equation}
M\ddot{x}=-\kappa x,
\end{equation}
with the solution%
\begin{equation}
x\left(  t\right)  =x_{0}\cos\left[  \omega_{0}t+\phi\right]
\end{equation}
where the angular frequency is $\omega_{0}=\sqrt{\kappa/M}$ and $\phi$ is a
constant phase angle. \ 

\subsection{Electromagnetic Oscillator}

In the case of the electromagnetic oscillator, we are considering only a
one-dimensional oscillator consisting of a charge $e$ with mass $M$ confined
along the $x$-axis by two charges $q$ of the same sign as $e.$ \ Thus we
consider two charges $q$ placed along the $x$-axis at $x_{q\pm}=\pm a$ with
the charge $e$ in unstable equilibrium at the origin of coordinates. \ If the
charge $e$ is displaced from the origin to the point $x$, then the force on
the charge is
\begin{equation}
\mathbf{F}=\widehat{x}qe\left[  \frac{1}{\left(  -a-x\right)  ^{2}}-\frac
{1}{\left(  a-x\right)  ^{2}}\right]  \approx-\widehat{x}\frac{4qe}{a^{3}}x.
\label{emosc}%
\end{equation}
Thus for small displacements $x$, there is a linear restoring force on the
charge $e$. \ If the system were treated as a purely mechanical system, the
(angular) frequency of of oscillation would be $\omega_{0}=\sqrt{4qe/\left(
a^{3}M\right)  }$. \ Now electromagnetism is a relativistic theory. \ Our
electromagnetic oscillator can be regarded as relativistic to any degree of
approximation provided that the amplitude $x_{0}$ of oscillation and hence the
maximum velocity $x_{0}\omega_{0}$ of the oscillation is taken as sufficiently
small, $x_{0}\omega_{0}<<c.$

\section{Equilibrium of Oscillators in Baths}

\subsection{Optional Mechanical Bath}

Nonrelativistic mechanics deals only with massive particles, and accordingly,
the bath with which the mechanical oscillator can be regarded as interacting
will be taken as composed of particles of mass $m$ moving in one dimension.
\ \ The oscillator particle of mass $M$ is connected by point collisions to a
bath of non-interacting particles of mass $m$ confined to a large
one-dimensional box of length $L$ with an elastically reflecting wall. \ The
distant end of the box provides the elastic-rebound wall, and the oscillator
particle $M$ (attached to the spring and wall) provides the other end of the
one-dimensional box. \ Then the point collisions of the bath particles with
the oscillator of mass $M$ provides a means of transfer of energy and momentum
between the bath particles and the oscillator system. \ 

Point collisions of massive particles lead to \textquotedblleft energy
sharing\textquotedblright\ among all the interacting particles of the system.
\ The work on kinetic theory carried out in the 19th century introduced the
idea of kinetic energy equipartition among the particles. \ Furthermore, the
mechanical harmonic oscillator of mass $M$ shares energy equally between its
kinetic energy and its potential energy. \ Thus the \textquotedblleft
energy-sharing\textquotedblright\ idea extends to all the modes of the
nonrelativistic mechanical oscillator coupled by point collisions to the bath
of non-interacting particles. \ This energy-sharing idea is carried over into
classical statistical mechanics and provides the basis for the Boltzmann
distribution on phase space. \ The system can be described satisfactorily by
classical statistical mechanics with the resulting Boltzmann probability
distribution. \ We note that the energy-sharing concept has no particular role
for the frequency of any oscillator nor any limit on the (finite) number of
parameters which may enter the nonrelativistic mechanical system interacting
with a finite number-density of particles. \ Equilibrium will involve the
preferred inertial frame of the confining box and the average kinetic energy
of a particle.

\subsection{Required Electromagnetic Bath}

For the electromagnetic oscillator, the bath is totally different from that of
the nonrelativistic mechanical situation. \ The mechanical oscillator can be
imagined to oscillate without friction and without any mechanical bath of
particles. \ In contrast, the charge $e$ of the electromagnetic oscillator may
oscillate harmonically without mechanical friction, but the oscillating charge
must accelerate. And the accelerating oscillaor charge emits radiation at its
natural frequency of oscillation $\omega_{0}.$ \ According to classical
theory, the oscillating electromagnetic particle is always coupled to
radiation. \ Thus within classical theory, equilibrium for an electromagnetic
oscillator \textit{requires} the presence of a bath of electromagnetic radiation.

Furthermore, electromagnetism is a relativistic theory. \ Although a bath of
mechanical particles with finite particle density can be described in terms of
a finite number of wave modes, the number of normal modes for relativistic
waves is infinite. \ Thus an electromagnetic oscillator is necessarily
connected to electromagnetic radiation involving an infinite number of wave
modes. \ It is natural to ask, \textquotedblleft What is the spectrum of the
radiation bath which is tied to a small (relativistic) motion of the charge
$e$ of the one-dimensional electromagnetic oscillator?\textquotedblright

\section{ Review of the Electromagnetic Radiation Bath in Dipole
Approximation}

\subsection{Random Classical Radiation}

An oscillator in a closed container with perfectly reflecting walls might be
in equilibrium with coherent radiation. \ The far more usual case is for the
oscillator to come to equilibrium with random radiation. \ The problem of an
electromagnetic oscillator interacting with random radiation is an old problem
going back to Planck's work at the end of the 19th century.\cite{Planck}
\ Here we will first review the historical dipole approximation calculation
before turning to the new aspects which are unfamiliar to most physicists. \ 

Random classical radiation can be described as Planck described it at then end
of the 19th century in terms of plane waves with random phases. \ If we
consider a very large cubic box with sides of length $a$, then the random
radiation can be written as%
\begin{equation}
\mathbf{E}(\mathbf{r},t)=%
{\displaystyle\sum_{\mathbf{k,\lambda}}}
\widehat{\epsilon}(\mathbf{k},\lambda)\left(  \frac{8\pi U_{rad}(\omega
)}{a^{3}}\right)  ^{1/2}\left\{  \exp\left[  i\mathbf{k}\cdot\mathbf{r}%
-i\omega t+i\theta\left(  \mathbf{k},\lambda\right)  \right]  +cc\right\}
\label{Eran}%
\end{equation}%
\begin{equation}
\mathbf{B}(\mathbf{r},t)=%
{\displaystyle\sum_{\mathbf{k,\lambda}}}
\widehat{\mathbf{k}}\times\widehat{\epsilon}(\mathbf{k},\lambda)\left(
\frac{8\pi U_{rad}(\omega)}{a^{3}}\right)  ^{1/2}\left\{  \exp\left[
i\mathbf{k}\cdot\mathbf{r}-i\omega t+i\theta\left(  \mathbf{k},\lambda\right)
\right]  +cc\right\}
\end{equation}
where the sum over the wave vectors $\mathbf{k}=\widehat{x}2\pi
l/a+\widehat{y}2\pi m/a+\widehat{z}2\pi n/a$ involves integers $l,m,n=0,\pm
1,\pm2,...$ running over all positive and negative values, there are two
polarizations $\lambda=1,2,$ and the random phases $\theta(\mathbf{k,\lambda
})$ are distributed uniformly over the interval $(0,2\pi],$ independently for
each wave vector $\mathbf{k}$ and polarization $\lambda$. \ The energy per
normal mode at radiation frequency $\omega$ is given by $U_{rad}(\omega),$ and
we have assumed that the radiation spectrum is isotropic.

\subsection{Oscillator Equation of Motion}

The one-dimensional electromagnetic oscillator is located at the origin and
oscillates along the $x$-axis. \ For small displacements $x$, the oscillator
motion satisfies Newton's second law%
\begin{equation}
M\ddot{x}=-M\omega_{0}x+M\tau\dddot{x}+eE_{x}(x,0,0,t),\label{Newton}%
\end{equation}
where the time $\tau=2e^{2}/(3Mc^{3}),$ and the $x$-component of the electric
field follows from Eq. (\ref{Eran}). \ In the electric dipole approximation
corresponding to a point oscillator, the electric field $E_{x}$ is
approximated as that located at the origin (center of the oscillator),
$E_{x}(x,0,0,t)\approx E_{x}(0,0,0,t).$ \ Substituting from Eq. (\ref{Eran}),
we have a linear stochastic differential equation with the steady-state
solution\cite{Uosc}%
\begin{equation}
x\left(  t\right)  =\frac{e}{M}%
{\displaystyle\sum_{\mathbf{k,\lambda}}}
\epsilon_{x}(\mathbf{k},\lambda)\left(  \frac{8\pi U_{rad}(\omega)}{a^{3}%
}\right)  ^{1/2}\frac{1}{2}\left\{  \frac{\exp\left\{  i\left[  \mathbf{k}%
\cdot\mathbf{r}-\omega t+\theta\left(  \mathbf{k},\lambda\right)  \right]
\right\}  }{-\omega^{2}+\omega_{0}^{2}+i\tau\omega^{3}}+cc\right\}
\label{xoft}%
\end{equation}
The time derivative $\dot{x}\left(  t\right)  $ follows from Eq. (\ref{xoft}).
\ The average values can be obtained by averaging over the random phases as
\begin{equation}
\left\langle \exp[i\theta(\mathbf{k},\lambda\mathbf{)]}\exp\left[
-i\theta\left(  \mathbf{k}^{\prime},\lambda^{\prime}\right)  \right]
\right\rangle =\delta_{\mathbf{kk}^{\prime}}\delta_{\lambda\lambda^{\prime}%
}.\label{angav1}%
\end{equation}
Thus averaging over the random phases and then summing over the Kronecker
delta, the mean-square displacement is%
\begin{equation}
\left\langle x^{2}\left(  t\right)  \right\rangle =%
{\displaystyle\sum_{\mathbf{k,\lambda}}}
\epsilon_{x}^{2}(\mathbf{k},\lambda)\left(  \frac{8\pi U_{rad}(\omega)}{a^{3}%
}\right)  \frac{e^{2}}{2m^{2}\left[  \left(  -\omega^{2}+\omega_{0}%
^{2}\right)  ^{2}+\left(  \tau\omega^{3}\right)  ^{2}\right]  }.
\end{equation}
Assuming that the box for the periodic boundary conditions becomes very large,
the sum over the discrete values of $\mathbf{k}$ can be replaced by an
integral, $%
{\textstyle\sum\nolimits_{\mathbf{k}}}
\rightarrow%
{\textstyle\int}
d^{3}k\left[  a/\left(  2\pi\right)  \right]  ^{3}=%
{\textstyle\int\nolimits_{0}^{\infty}}
k^{2}dk%
{\textstyle\int}
d\Omega\left[  a/\left(  2\pi\right)  \right]  ^{3}.$ \ The only angular
dependence appears in the polarization vectors $\epsilon_{x}^{2}%
(\mathbf{k},\lambda)$, so that, on angular integration, each polarization will
contribute a value of $1/3.$ \ We are left with the integration over $k.$ \ If
we assume that the charge $e$ is small so that the damping is small and the
integrand is sharply peaked at $\omega_{0}$, then we may extend the lower
limit of the integral to $-\infty$, and replace all frequencies $\omega$ by
$\omega_{0},$ except where the combination $\omega-\omega_{0}$ appears. \ The
remaining integral is of the form%
\begin{equation}%
{\displaystyle\int_{-\infty}^{\infty}}
\frac{du}{a^{2}u^{2}+b^{2}}=\frac{\pi}{ab}.
\end{equation}
The mean-square displacement is%
\begin{equation}
\left\langle x^{2}\right\rangle =\frac{U_{rad}(\omega_{0})}{m\omega_{0}^{2}}.
\end{equation}
An analogous procedure can be followed for the evaluation of $\left\langle
\dot{x}^{2}\right\rangle .$ Then the average energy of the oscillator follows
as%
\begin{equation}
U_{osc}\mathcal{=}\left\langle \frac{1}{2}M\dot{x}^{2}+\frac{1}{2}M\omega
_{0}^{2}x^{2}\right\rangle =U_{rad}\left(  \omega_{0}\right)  ,
\end{equation}
so that in equilibrium (for the small-charge approximation) the point dipole
oscillator has the same average energy as the radiation modes at the same
frequency as the oscillator.\cite{Uosc}

\subsection{Phase Space Distribution for the Electromagnetic Oscillator}

One can also obtain the averages for all products involving an arbitrary
number of factors of $x\left(  t\right)  $ and $\dot{x}\left(  t\right)  $ by
repeated use of the average in Eq. (\ref{angav1}).\cite{gen} \ Indeed, the
averages correspond to a probability distribution $P(x,p)$\ for the
displacement $x\left(  t\right)  $ and momentum $p\left(  t\right)  =M\dot
{x}\left(  t\right)  ,$ for $-\infty<x<\infty,$ $-\infty<p<\infty,$ given
by\cite{Uosc}\cite{gen}
\begin{equation}
P_{osc}(x,p)dxdp=\frac{1}{2\pi U_{osc}}\sqrt{\frac{\kappa}{M}}\exp\left[
-\frac{\left(  \kappa x^{2}+p^{2}/M\right)  }{2U_{osc}}\right]  dxdp.
\end{equation}

It is interesting to rewrite the probability distribution for the oscillator
in random radiation in terms of action-angle variables\cite{Goldstein} using
\begin{align}
x\left(  t\right)   &  =\sqrt{\frac{2J}{m\omega_{0}}}\cos w\nonumber\\
p\left(  t\right)   &  =\sqrt{2mJ\omega_{0}}\cos w
\end{align}
Then the phase space distribution for the electromagnetic oscillator, for
$0\leq w\leq2\pi,0\leq J<\infty,$ is given by
\begin{equation}
P^{osc}(w,J)dwdJ=\frac{1}{2\pi}\frac{U_{osc}}{\omega_{0}}\exp\left[
-\frac{J\omega_{0}}{U_{osc}}\right]  dwdJ. \label{Posc}%
\end{equation}

As shown in earlier work,\cite{appen} the scatting of the random radiation by
the point electric dipole oscillator does not change the frequency spectrum or
the isotropic nature of the random radiation. \ We emphasize that the energy
$U_{osc}$ of the point electric dipole oscillator matches the radiation energy
$U_{rad}$ in the normal modes of the radiation field at the oscillator natural
frequency $\omega_{0}$. \ However, this is merely a connection at one point
$\omega=\omega_{0}$ of the spectral function $U_{rad}(\omega)$. \ There is
nothing in this calculation which suggests any preferred spectrum for the
random radiation in equilibrium with the electromagnetic oscillator in dipole approximation.\ \ 

\subsection{Historical Use of the Electromagnetic Oscillator in Dipole
Approximation}

It was Planck who introduced the electromagnetic oscillator taken in the
dipole approximation in connection with the problem of blackbody radiation at
the end of the 19th century.\cite{Planck} \ The oscillator was regarded as a
connection between the thermodynamics of matter and the corresponding
electromagnetic thermal radiation. \ Initially, Planck hoped that the
scattering of radiation by the oscillator itself would force electromagnetic
radiation into equilibrium and so reveal the blackbody radiation spectrum.
\ However, Boltzmann pointed out that electromagnetic theory is invariant
under time reversal so that Planck's hope was empty.\cite{history} \ It was
only when Planck became convinced that the dipole oscillator would not change
the frequencies of scattered radiation and so would not push a spectrum of
random radiation towards equilibrium that Planck turned to statistical
mechanical ideas for particles. \ Only then did Planck seek to obtain the
thermodynamic behavior of the electromagnetic oscillator in order to determine
the equilibrium spectrum of the associated thermal radiation.\cite{history}%
\cite{blackbody}

The historical use of the electromagnetic dipole oscillator still appears in
the textbooks of modern physics.\cite{mod} \ Using the ideas of
nonrelativistic classical statistical mechanics for the oscillator, the texts
claim that a classical electromagnetic oscillator must assume the
equipartition energy $U_{osc}=k_{B}T$ of a Newtonian mechanical oscillator,
and therefore, by Planck's calculation showing the equality of energy between
the electromagnetic dipole oscillator and the radiation normal modes, the
spectrum of thermal radiation $U_{rad}$ must be the Rayleigh-Jeans spectrum
\begin{equation}
U_{rad}\left(  \omega,T\right)  =U_{RJ}(\omega,T)=k_{B}T. \label{R-J}%
\end{equation}
Although energy equipartition is strictly a result of nonrelativistic
mechanics and has nothing to do with relativistic classical electromagnetism,
it is claimed that this Rayleigh-Jeans result is the prediction of
\textquotedblleft classical physics.\textquotedblright\ \ And this prediction
is made without any allowance for the conflict between nonrelativistic
mechanics and relativistic electromagnetism.

This same result that nonrelativistic physics leads to the Rayleigh-Jeans
radiation spectrum has been obtained by other researchers without using
classical statistical mechanics but by using nonrelativistic nonlinear
scattering systems which are connected to electromagnetic radiation by the
dipole radiation approximation for the mechanical systems.\cite{nonlin}
\ Whereas the analysis of the modern physics texts uses nonrelativistic
classical statistical mechanics, all of these scattering calculation use
nonrelativistic nonlinear potential systems for the radiation scatterers.
\ All of these derivations give a false result because\ (despite any claims to
being relativistic calculations) they all exclude Coulomb potentials and
involve a nonrelativistic basis for the attainment of equilibrium. \ 

\section{Relativistic Radiation Equilibrium}

\subsection{Few Relativistic Scattering Systems}

The use of nonrelativistic mechanical systems in connection with
electromagnetic radiation equilibrium persists because there are very few
relativistic scattering systems in nature which are familiar to physicists and
which allow tractable analytic calculations. \ Indeed, only the system of a
point charge in a Coulomb potential when coupled to electromagnetic radiation
can be regarded as fully relativistic. \ And analytic treatment of the
Coulomb-potential system seems extraordinarily difficult. \ To date, only
numerical simulations have provided some insight into the Coulomb system in
radiation, and then only in the nonrelativistic approximation for the particle
motion.\cite{Cole} \ The one other system which can be regarded as
approximately relativistic is that of a point charge in a one-dimensional
harmonic potential when the amplitude of oscillation is taken as very small so
that the particle velocity is very small, $v<<c$.

\subsection{Analysis of the Electromagnetic Oscillator Beyond the Dipole
Approximation}

Although the classical electromagnetic calculation for the point electric
dipole oscillator in random radiation has been presented repeatedly for a
century, the possibility of going beyond the dipole approximation was
emphasized only recently in work by Huang and Batelaan\cite{HB} who considered
the absorption of a radiation pulse by an oscillator of non-zero amplitude.
\ We will not consider the absorption of a radiation pulse, but, inspired by
the work of Huang and Batelaan, we wish to continue the analysis of the
equilibrium between an electromagnetic harmonic\ oscillator and a radiation
bath by considering the higher harmonics. \ We note that by taking the
amplitude of oscillation as small enough, the motion of the oscillating charge
$e$ can be regarded as relativistic to any degree of approximation, $v<<c,$
and yet, provided that the amplitude of motion is non-zero, there is always
non-zero radiation at harmonics. \ 

Thus if the amplitude of oscillation of the electromagnetic oscillator is not
zero, then the harmonic oscillator of natural frequency $\omega_{0}$ must be
emitting quadrupole radiation at frequency $2\omega_{0}.$ \ But here indeed is
an opportunity to determine the equilibrium spectrum of a very small
electromagnetic oscillator. \ There must be energy in the random radiation
spectrum at frequency $2\omega_{0}$ which is absorbed by the oscillator and
balances the energy radiated as quadrupole radiation at $2\omega_{0}.$ \ And
indeed, the argument can be repeated for all the higher radiation multipoles
of the oscillating charge $e.$ \ The situation here is totally different from
the earlier scattering calculations\cite{nonlin}, all of which involved
nonrelativistic non-linear oscillators scattering electromagnetic radiation in
the dipole approximation, and all of which led to the Rayleigh-Jeans spectrum
as the equilibrium spectrum of random radiation. Here for an electromagnetic
oscillator in a purely harmonic potential, the equilibrium spectrum for the
electromagnetic oscillator is not determined by any assumption about the
mechanical structure of the oscillator, but rather the spectrum is determined
by purely electromagnetic considerations. \ This is the first scattering
calculation for electromagnetic radiation which can be regarded as fully
relativistic.\cite{Equ}

The change from the earlier work involves the equation of motion for the
oscillator given in Eq. (\ref{Newton}), but now \textit{not} taking the dipole
approximation $E_{x}(x,0,0,t)\approx E_{x}(0,0,0,t),$but rather, for the
quadrupole term, making the next approximation%
\begin{equation}
E_{x}(x,0,0,t)\approx E_{x}(0,0,0,t)+x(t)\left(  \frac{d}{dx^{\prime}}%
E_{x}(x^{\prime},0,0,t)\right)  _{x^{\prime}=0}%
\end{equation}
where on the right-hand side we will insert for $x\left(  t\right)  $ the
dipole-approximation result appearing in Eq. (\ref{xoft}). \ Indeed, the whole
idea is to make successive approximations in the small amplitude of oscillator
motion. \ Thus now for the quadrupole radiation approximation, we write the
equation of motion for the oscillating charge $e$ as%
\begin{equation}
M\ddot{x}\approx-M\omega_{0}x^{2}+M\tau\dddot{x}+eE_{x}(0,0,0,t)+x^{\left(
0\right)  }\left(  t\right)  \left(  \frac{d}{dx^{\prime}}E_{x}(x^{\prime
},0,0,t)\right)  _{x^{\prime}=0} \label{Nx2}%
\end{equation}
where $x^{\left(  0\right)  }\left(  t\right)  $ is the dipole approximation
appearing in Eq. (\ref{xoft}). \ We notice that whereas the equation of motion
for the electromagnetic oscillator in the dipole approximation involved one
factor of $\exp\left[  -i\omega t+i\theta(\mathbf{k,\lambda)}\right]  $ on the
right-hand side there are now two factors of this form in Eq. (\ref{Nx2}).

One can solve\cite{Equ} the equation of motion (\ref{Nx2}), and one finds that
the scattered radiation preserves the frequency spectrum and the angular
distribution of isotropic radiation provided that the spectrum of random
radiation has twice as much energy per normal mode at frequency $2\omega_{0}$
compared to the energy per normal mode at frequency $\omega_{0}.$\cite{Equ}
\ It is easy to see where this analysis is going. \ The electromagnetic
oscillator in one dimension is in equilibrium only with a radiation bath
$U_{rad}(n\omega_{0})=const\times n\omega_{0}$ for all multiples $n$ of the
fundamental oscillator frequency $\omega_{0}.$Thus detailed balance for
radiation energy holds provided that the spectrum of random radiation is
linear in frequency%
\begin{equation}
U_{rad}(\omega)=const\times\omega.
\end{equation}
This situation corresponds to Lorentz invariance for the spectrum.\cite{rel}
\ Indeed, it seems natural that the Lorentz-invariant theory of classical
electrodynamics should pick out a Lorentz-invariant spectrum of random
radiation for the electromagnetic oscillator which has no structure other than
the characteristic frequency $\omega_{0}$.

\section{Adiabatic Change and Temperature}

\subsection{Adiabatic Change of the Mechanical Spring Constant}

Adiabatic changes in the oscillator frequency for a Newtonian mechanical
oscillator in equilibrium through point collisions with a particle bath show
that the temperature of the bath can never be zero unless the total system
energy vanishes. \ If the spring constant $\kappa$ of the oscillator is
readjusted very slowly, then the external agent which provides the
readjustment is carrying out an adiabatic change of the system. \ If we
imagine that the collision interactions between the oscillator and the bath
particles are removed during the adiabatic change of the oscillator, then the
ratio of the energy $U_{osc}$ to the oscillator (angular) frequency
$\omega_{0}$ is an adiabatic invariant, $U_{osc}/\omega_{0}=J_{osc},$ where
$J_{osc}$ is the action variable\cite{Goldstein531} in the action-angle
treatment of the mechanics of the oscillator, with $U_{osc}=J_{osc}\omega_{0}%
$. \ Thus during the adiabatic readjustment of the spring constant, the
probability distribution of the action variable $P\left(  J_{osc}\right)  $ is
unchanged, but the average energy of the oscillator does indeed change since
$U_{osc}=J_{osc}\omega_{0},$ and $\omega_{0}$ has changed.

If the mechanical oscillator is now reconnected to the bath of particles, the
oscillator will no longer be in equilibrium with the bath because its average
energy per normal mode has been changed. \ Rather, energy will be exchanged
between the oscillator and the bath by means of point collisions, and a new
equilibrium will be established with a different total energy for the combined
system of the oscillator and the bath. \ Since the adiabatic change is not the
same as a isothermal change, the temperature of the oscillator and its
particle bath must be regarded as non-zero. \ Thus provided that the
mechanical motion of the oscillator and its surrounding bath particles has not
actually ceased, the oscillator-bath system must be regarded as at equilibrium
at some positive temperature determined by the average kinetic energy of any
of the particles. \ There is no such thing as particle motion at the absolute
zero of temperature in nonrelativistic classical mechanics. \ 

\subsection{Adiabatic Change of the Frequency of the Electromagnetic
Oscillator}

In contrast with the Newtonian mechanical oscillator, an adiabatic change in
the oscillator frequency of a one-dimensional electromagnetic oscillator in
equilibrium with random radiation shows that the
one-dimensional-electromagnetic-oscillator-radiation system has zero
temperature. \ 

Electromagnetism depends crucially on frequency-dependent resonances whereas
point collisions do not have any associated frequency. \ Thus the frequency of
the Newtonian mechanical oscillator is largely irrelevant in connection with
the particle bath. \ However, the frequency of the electromagnetic oscillator
determines the radiation frequencies with which the electromagnetic oscillator
will interact. \ If we imagine the electromagnetic oscillator temporarily
uncoupled from its radiation and the separation $a$ between the confining
charges $q$ as changed, then the adiabatic change of the oscillation frequency
$\omega_{0}$ appearing in Eq. (\ref{emosc}) leads to a change in energy
$\Delta U_{osc}=J_{osc}\Delta\omega_{0}$ of the oscillator while keeping the
value of the action variable $J_{osc}$ fixed. \ For an ensemble of
oscillators, the phase space distribution involving $J_{osc}$ is unchanged
under the adiabatic change in frequency, but the average oscillator energy is
still linear in frequency with the same phase space distribution as given in
Eq. (\ref{Posc}) since the ratio of the oscillator energy to the oscillator
frequency has not changed. \ But now when the electromagnetic oscillator is
reconnected with the initial Lorentz-invariant spectrum of random radiation,
there is no transfer of average energy between the oscillator and the bath of
electromagnetic radiation.\cite{adiabem} \ The transformed oscillator is again
in equilibrium with the original radiation spectrum. \ \ Thus an adiabatic
change is the same as an isothermal change for this situation, and we would
describe the one-dimensional electromagnetic oscillator as in equilibrium with
its radiation bath at the absolute zero of temperature, $T=0.$ \ 

We emphasize that the energy introduced in the adiabatic change went into the
energy of the oscillator, and (on average) no energy was transferred to the
energy-divergent radiation field. \ There is \textquotedblleft apparent
decoupling\textquotedblright\ between the electromagnetic oscillator and its
equilibrium radiation bath. \ Thus in a sense the electromagnetic oscillator
acts as though it were decoupled from the radiation field even though the
radiation field is required for the stability of the oscillating charge in
classical electromagnetism. \ In this connection, it is noteworthy that
textbooks of thermodynamics and statistical mechanics will often introduce
oscillators and electromagnetic systems without worrying about any divergent
radiation bath which would be required for their stability within classical
physics. \ \ 

\section{Classical Electromagnetic Zero-Point Radiation}

\subsection{Unique Lorentz-Invariant Spectrum}

In nature, oscillating charges can not each have their own Lorentz-invariant
spectrum of random radiation. \ Thus classical electromagnetic theory demands
that in order to have oscillating electric charges, there must be one
Lorentz-invariant spectrum of random classical electromagnetic radiation.
\ This radiation spectrum is usually termed classical electromagnetic
zero-point radiation.\cite{under} \ And within a purely classical
electromagnetic description of nature, there is good experimental evidence for
classical electromagnetic zero-point radiation with an energy per radiation
normal mode%
\begin{equation}
U_{rad}\left(  \omega\right)  =\left(  1/2\right)  \hbar\omega,
\end{equation}
where the constant $\hbar$ takes the same value as is given for Planck's
constant $\hbar=h/\left(  2\pi\right)  .$ \ 

The existence of a spectrum of classical electromagnetic zero-point radiation
provides the basis for classical explanations of Casimir forces, van der Waals
forces, the specific heats of solids, diamagnetism, and the absence of atomic
collapse.\cite{review}\ 

\subsection{Phase Space of Classical Zero-Point Radiation}

An electromagnetic oscillator in zero-point radiation takes on the same phase
space distribution as is found for the radiation normal modes. Both the
electromagnetic oscillator and all the modes of the electromagnetic radiation
field have the distribution on phase space following from Eq. (\ref{Posc}) for
$0\leq w\leq2\pi,0\leq J<\infty,$%
\begin{equation}
P(w,J)dwdJ=\frac{1}{2\pi}\frac{\hbar}{2}\exp\left[  -\frac{J}{\hbar/2}\right]
dwdJ,
\end{equation}
which does not depend upon the frequency of the oscillator or of the radiation
mode.\cite{PJ}\cite{adiabem} \ On adiabatic change of the frequency of a
purely-harmonic oscillator, the phase space distribution is unchanged and
remains in equilibrium with the zero-point radiation spectrum.

\subsection{Spectrum of \textquotedblleft Least Possible
Information\textquotedblright}

The Lorentz-invariant spectrum of zero-point radiation is the
\textquotedblleft radiation spectrum of least possible information in a
relativistic theory.\textquotedblright\ \ The zero-point radiation spectrum
involves one overall constant $\hbar$ and takes the same isotropic form in
every inertial frame. \ On the other hand, the Rayleigh-Jeans spectrum
contains more information than the Lorentz-invariant zero-point radiation
spectrum. \ Thus in addition to one over-all parameter $k_{B}T$, the
Rayleigh-Jeans spectrum has a preferred inertial frame in which the simple
Rayleigh-Jeans form in Eq. (\ref{R-J}) appears; any other inertial frame
moving with constant velocity relative to the preferred frame will find a
non-isotropic spectrum of radiation. \ Nonrelativistic nonlinear oscillators
scatter radiation so as to impose their own inertial frame as the preferred
inertial frame of the random radiation spectrum to which they are attached.
\ In contrast, the calculation\cite{Equ} showing that the one-dimensional
electromagnetic oscillator in a purely-harmonic potential is in equilibrium
with a Lorentz-invariant radiation spectrum indeed suggests that the
purely-harmonic oscillator is not imposing its own inertial frame on the
equilibrium radiation spectrum upon which it depends. \ Indeed, the
electromagnetic oscillator based upon the electrostatic consideration leading
up to Eq. (\ref{emosc}) will have \textit{relativistic} energy transformation
properties between inertial frames provided the the particle velocity is small
in the oscillator frame. \ On the other hand, the potential energy of a
nonrelativistic nonlinear oscillator will not transform in a relativistic fashion.

In addition to providing a Lorentz-invariant radiation spectrum in Minkowski
spacetime, zero-point radiation allows incorporation into general relativity
where it again corresponds to the \textquotedblleft radiation spectrum of
least possible information in a relativistic theory.\textquotedblright\ \ In
the general relativistic situation, the correlation functions for zero-point
radiation between spacetime points depend upon only the geodesic separations
between the spacetime points.\cite{geo}\ 

\section{Limitations of the Analysis}

The one-dimensional electromagnetic oscillator in a purely-harmonic potential
scatters radiation so as to push the radiation toward its equilibrium spectrum
which is a Lorentz-invariant spectrum.\cite{Equ} \ This is indeed consistent
with classical electrodynamics which is a Lorentz-invariant theory. \ However,
zero-point radiation must be interpreted as corresponding to a temperature of
abslute zero, and the scattering analysis does not suggest any equilibrium at
a thermal spectrum with non-zero temperature. \ Also, we note that the
one-dimensional oscillator shows no velocity-dependent damping, which must
exist for thermal radiation at non-zero temperature. Indeed the
one-dimensional analysis allows no role for the magnetic field of the random
radiation which would cause forces perpendicular to the direction of
oscillation which are eliminated by the constraint giving one-dimensional
motion. \ Finally, the one-dimensional oscillator is not a purely
electromagnetic system because of the non-electromagnetic constraint which
gives one-dimensional motion. \ Indeed, Earnshaw's theory tells us that a
stable purely-harmonic potential cannot be obtained from interacting point
charges. \ 

It is conjectured that all of these considerations are related. \ It is
suggested that velocity-dependent damping in random radiation is possible only
if the charged particle can oscillate (in a sort of zitterbewegung) in the
direction perpendicular to the direction of damping. \ Indeed, the
Einstein-Hopf analysis\cite{EH} for the motion of a classical particle in
thermal radiation involves a dipole oscillator, internal to the particle,
which oscillates in a direction perpendicular to the direction of the
particle's constrained one-dimension motion. \ Thus it is conjectured that the
complete understanding of the thermal radiation spectrum in terms of
scattering by a purely classical system must wait until there is a classical
understanding of something like the classical hydrogen atom. \ While awaiting
such a classical scattering analysis, one is encouraged by many points of view
that suggest that thermal radiation can indeed be understood in terms of
classical physics.\cite{blackbody}

\section{Concluding Comments}

The biggest shock to undergraduate physics students is their encounter with
the unusual ideas of quantum theory. \ However, they are often not confronted
with just how different the ideas of classical electrodynamics are from the
every-day experiences of Newtonian mechanics. \ Like the physicists of the
early 20th century, contemporary physicists still do not appreciated the
conflict between Newtonian mechanics and electromagnetism. \ Furthermore, the
textbooks of modern physics continue to mislead physics students regarding
this matter by insisting that the energy-sharing ideas of nonrelativistic
classical mechanics are to be carried over to the relativistic theory of
classical electromagnetic radiation equilibrium. \ 

Newtonian mechanics and classical electrodynamics are vastly different in
their assumptions and in their descriptions of nature. \ A nonrelativistic
mechanical harmonic oscillator can be assumed to have no friction, and so will
oscillate indefinitely with no need for any bath associated with the
equilibrium situation. \ In addition, nonrelativistic mechanical systems of
particles come to equilibrium by energy-sharing throughout the mechancal
system. \ This energy-sharing idea appears in the equipartition theorem and in
the Boltzmann distribution which are incorporated into nonrelativistic
classical statistical mechanics. \ 

In contrast with the mechanical situation, an electric charge is always
connected to electromagnetic fields. \ If the electric charge oscillates, then
the charge must exchange energy with the radiation field. \ There is no such
thing as a classical electromagnetic oscillator which is not coupled to
electromagnetic radiation. \ Furthermore, a relativistic field theory will
involve an infinite number of radiation normal modes. \ Thus the
nonrelativistic idea of energy-sharing across the entire system has no place
in electromagnetism. \ Rather, resonance phenomena associated with frequencies
appear crucially in electromagnetism, whereas such frequency-dependent
resonances are of no significance in nonrelativistic mechanical equilibrium. \ 

It is very hard to find relativistic electromagnetic systems allowing easy
calculations involving radiation equilibrium. \ In this article, we have
considered the simplest relativistic classical electromagnetic system which
will lead to a preferred equilibrium radiation spectrum. \ The oscillation of
an electric charge in a one-dimensional purely-harmonic potential leads to a
required equilibrium radiation spectrum, not through its mechancal motion
which is always simply harmonic, but through the electromagnetic aspects of
the multipole radiation which arise in classical electromagnetism. \ One finds
that the one-dimensional electromagnetic oscillator is in equilibrium with a
spectrum whose normal mode energy $U_{rad}$ is linear in frequency $\omega$,
$U_{rad}=const\times\omega,$ corresponding to a Lorentz-invariant radiation
spectrum. \ Indeed, the relativistic nature of the equilibrium spectrum fits
with the relativistic nature of classical electromagnetism. \ \ 

\section{Acknowledgement}

I am deeply indebted to the work of Dr. Wayne Huang and Dr. Herman Batelaan
and of Dr. Daniel C. Cole who have emphasized the importance of radiation
harmonics in their analyses of electromagnetic systems. \

\end{document}